\documentclass[prd,preprint]{revtex4}   

\begin{document}

\title{{\bf Loop Quantum Gravity Corrections and Cosmic Ray Decays}}

\author{Jorge Alfaro and  Gonzalo Palma}

\affiliation{Facultad de F\'{\i}sica, Pontificia Universidad Cat\'{o}lica de Chile \\
        Casilla 306, Santiago 22, \textsc{Chile}.
\\ {\tt jalfaro@puc.cl, \  gpalma@astro.puc.cl}}

\date{\today}


\begin{abstract}
Loop quantum gravity effective theories are reviewed in the
context of the observed GZK limit anomaly and related processes.
This is accomplished through a kinematical analysis of the
modified threshold conditions for the involved decay reactions,
arising from the theory. Specially interesting is the possibility
of an helicity dependant violation of the limit, whose primary
effect would be the observation of favoured helicity states for
highly energetic particles.
\end{abstract}

\maketitle

\section{Introduction}

The vast void that still separates us from a definite version of a
quantum theory of gravity, and the fact that several alleged
versions of it are being proposed, has motivated the development
of various semiclassical approaches. These approaches follow the
form of effective theories which take into consideration
matter-gravity couplings, such as exposed in a number of recent
works \cite{Amelino & Ellis, Gambini Pullin, Neutrinos, Photons},
whose main results are the introduction of new terms in the
equations of motion for the described system. An inevitable
outcome of these works is the introduction of Lorentz Invariance
Deformations (LID) at the effective theory level. These
deformations become manifest when one analyzes the dispersion
relations for freely propagating particles, and may have notorious
consequences in high energy phenomena.

In particular, both \cite{Neutrinos} and \cite{Photons} are based
on the Loop Quantum Gravity (LQG) framework \cite{LQG}. In these
works, the effects of the loop structure of space, at the Planck
level, are treated semiclassically through a coarse-grained
approximation. An interesting feature of this kind of methods is
the appearance of a new length scale $\mathcal{L}$ (with
$\mathcal{L} \gg$ Planck length $\ell_{p}$), such that for
distances $d \ll \mathcal{L}$ the quantum loop structure of space
is manifest, while for distances $d \geq \mathcal{L}$ the
continuous flat geometry is regained. This scale gives us the hope
of bringing the effects of quantum gravity at an observable level.
A natural question thus arise. ¿Are we actually observing these
quantum gravity effects?. To answer this question we are forced to
go through the observations of the greatest energy registered.

The most energetic measured events are found in the form of Ultra
High Energy Cosmic Rays (UHECR) \cite{exp1, exp2}. Such events
(energies above $10^{20} eV$) are actually violating the
theoretical threshold known as the GZK limit \cite{Greisen,
Zatsepin & Kuz'min}, from which no extragalactic cosmic ray can
exceed, in energy, the value of $5 \times 10^{19} eV$. This
current limit takes into consideration the interaction of protons
with photons from the Cosmic Microwave Background Radiation
(CMBR). There have been different attempts to formulate a
convincing explanation about why such energetic particles are
reaching the Earth. In a purely theoretical fashion, perhaps the
most interesting explanations are the manifestation of decaying
magnetic monopoles \cite{Monopoles}, and the decay of super-heavy
relic particles \cite{Super-relic}. Another more orthodox
explanation can be found in the existence of Z-burst produced by
collisions between ultra high energy neutrinos and cosmic relic
neutrinos \cite{Z-Burst1, Z-Burst2, Z-Burst3}. However, neither of these
previous possibilities are fully satisfactory.

Another relevant observation is the detection of extragalactic
multi-TeV photons from the BL Lac object known as Markarian (Mrk)
501 \cite{Markarian}. These detected photons have reached energies
up to $20 TeV$. Similar to the case of protons, these multi-TeV
$\gamma$-rays are subject to the interaction with the Far Infrared
Background Radiation (FIBR), setting a limit to the energy of the
photons that can reach us. Initially, the collected data suggested
a violation of this limit, though it has recently been stated that
no such violation exists \cite{Stecker & Glashow, Stecker &
Jager}. We adopt this last position.

In this paper we study the possible bounds on the length scale
$\mathcal{L}$ emerging from the two observations mentioned above.
This is accomplished through a kinematical analysis of the
threshold conditions for the decays to be possible. In particular,
since the GZK limit is broken, we assume that a reasonable
explanation is found in the LID offered by the theory (see
\cite{Coleman & Glashow, Amelino, Stecker & Glashow} for other
similar approaches). On that score, the LID manifestations will,
in certain cases, depend on the difference between two LQG
parameters, each one belonging to different particles. For
instance, as shown in \cite{Coleman & Glashow}, if the dispersion
relation for a particle $i$ is (from here on, $\hbar = c = 1$)
\begin{eqnarray}
E_{i}^{2} = A_{i}^{2} p_{i}^{2} + m_{i}^{2}
\end{eqnarray}
(where $E_{i}$, $p_{i}$ and $m_{i}$ are the respective energy,
momentum and mass of the $i$th particle, and $A_{i}$ is a LID
parameter that can be interpreted as the maximum velocity of the
$i$th particle) then, one can show that the mentioned thresholds
can be substantially modified provided that the difference $\delta
A = A_{a} - A_{b}$ is not null ($a$ and $b$ are two particles
involved in the reaction leading to the mentioned threshold). Of
course, this effect compromises the universality of the given
parameters, namely, the fact that the $A_{i}$ parameters
---which eventually contain the information regarding the matter-gravity coupling---
were not the same for all particles. In the case of the current
LQG effective theories, these non universal deviations could be
understood as the manifestation of the breakup of classical
symmetries, emerging as a consequence of the choice of the quantum
gravity vacuum. In this way, the standard model structure of
different particles could appear through differentiated values for
the parameters in question. In this respect, since we do not have
a detailed knowledge of the precise values of the correction
parameters, we shall consider all possible scenarios for the
mentioned observations.

Finally we must mention the fact that in general, the presence of
LID forces us to consider the appearance of a preferred reference
system. In the case of the LQG corrections that we will consider,
the dispersion relations are valid only in an isotropic system.
For this reason, we shall naturally assume that this preferred
system is the CMBR co-moving reference frame and consequently, the
threshold conditions for the different decays should be considered
keeping this in mind.

\section{Dispersion Relations From Loop Quantum Gravity}

Here we present the main results from \cite{Neutrinos} and
\cite{Photons} relative to the modifications of the dispersion
relations of freely propagating neutrinos (more precisely Majorana
fermions) and photons. We shall assume that the results for
Majorana fermions are extensive to fermions in general. This
assumption relies on the fact that no substantial departure from
the original methods would be expected for the general case, since
the only difference is that for Majorana fermions one must impose
the reality condition on to the field equations. Of course, one
could expect that in the case of more general fermions there would
appear more corrective terms. Nevertheless, from the symmetry
arguments found in \cite{Neutrinos}, we should not expect new
$\mathcal{L}$ and $\ell_p$ dependant corrections different from
that which already appear in the present theory.

Special attention deserves the appearance of the length scale
$\mathcal{L}$.

\subsection{Fermions}

For Majorana fermions \cite{Neutrinos}, the dispersion relation is
given by
\begin{eqnarray} \label{eq: E fermion}
E_{\pm}^{2} = (Ap \pm \frac{B}{2 \mathcal{L}})^{2} + m^{2} (\alpha
\pm \beta p)^{2}
\end{eqnarray}
where
\begin{eqnarray}
A &=& \left( 1+ \kappa_{1} \frac{\ell_{p}}{\mathcal{L}}+\kappa_{2}
\left( \frac{\ell_{p}}{\mathcal{L}}  \right)^{2}
+\frac{\kappa_{3}}{2} \ell_{p}^{2} p^{2} \right), \nonumber \\
B &=& \left( \kappa_{5} \frac{\ell_{p}}{\mathcal{L}}+\kappa_{6}
\left( \frac{\ell_{p}}{\mathcal{L}}  \right)^{2}
+\frac{\kappa_{7}}{2} \ell_{p}^{2} p^{2} \right), \nonumber \\
\alpha &=& \left( 1+ \kappa_{8} \frac{\ell_{p}}{\mathcal{L}}  \right), \nonumber \\
\beta &=& \frac{\kappa_{9}}{2} \ell_{p}.
\end{eqnarray}
In the former expressions, $E_{\pm}$ is the energy of the
fermionic particle of mass $m$ and momentum $p$, and the
$\kappa_{i}$ are unknown adimensional parameters of order one. The
$\pm$ signs stand for the helicity of the propagating fermion. It
should be stressed that the terms associated with $B$ and $\beta$,
and which are precisely causing the $\pm$ signs, are both parity
and CPT odd (in fact, the equations of motion are invariant under
charge conjugation and time reversal operations).

In what follows, it will be sufficient to consider
\begin{eqnarray} \label{eq: E for fermions}
E^{2}_{\pm} = A^{2} p^{2} + \kappa_{3} \, \ell_{p}^{2} \, p^{4}
\pm \kappa_{5} \frac{\ell_{p}}{\mathcal{L}^{2}} |p| + m^{2} +
\frac{1}{4} \left( \kappa_{5} \frac{\ell_{p}}{\mathcal{L}^{2}}
\right)^{2},
\end{eqnarray}
where now $A = 1 + \kappa_{1} \, \ell_{p}/\mathcal{L}$ and
$\kappa_{1}$, $\kappa_{3}$ and $\kappa_{5}$ are of order one. For
simplicity, let us write (with $\eta = \kappa_{3} \ell_{p}^{2}$
and $\lambda = \kappa_{5} \ell_{p} / 2 \mathcal{L}^{2}$)
\begin{eqnarray} \label{eq: E for fermions 2}
E^{2}_{\pm} = A^{2} p^{2} + \eta p^{4} \pm 2 \lambda p + m^{2},
\end{eqnarray}
where we have absorbed the quadratic term in $\kappa_{5}$ in to
the mass. As we have said, the basis of the present work relies on
the assumption that (\ref{eq: E fermion}) is a valid expression
for fermionic particles in general. In particular, we will adopt
the expression (\ref{eq: E for fermions 2}) for electrons, protons
and $\Delta$ particles.

\subsection{Photons}

For photons \cite{Photons}, the dispersion relation is
\begin{eqnarray} \label{eq:complet E for photons}
E_{\pm} = p [A_{\gamma} - \theta_{3}(\ell_{p} p)^{2} \pm
\theta_{8} \ell_{p} p],
\end{eqnarray}
where
\begin{eqnarray}
A_{\gamma} &=& 1 + \kappa_{\gamma} \left(
\frac{\ell_{p}}{\mathcal{L}} \right)^{2+2\Upsilon}.
\end{eqnarray}
In the previous expressions, $E_{\pm}$ and $p$ are the respective
energy and momentum of the photon, while $\kappa_{\gamma}$ and
$\theta_{i}$ are adimensional parameters of order one, and
$\Upsilon$ is a free parameter that, for the moment, still needs
interpretation (it should be noted that the presence of the
$\Upsilon$ parameter in the fermion dispersion relation was not
considered in \cite{Neutrinos}). For simplicity we shall only
consider the possibilities $\Upsilon$ = -1/2, 0, 1/2, 1, etc... in
such a way that $A_{\gamma} \sim 1 +
\mathcal{O}[(\ell_{p}/\mathcal{L})^{n}]$, with $n=2+2\Upsilon$ a
positive integer. With this assumption, we will be able to find a
tentative value for $\Upsilon$, through the bounding of the lower
order correction of $\delta A \sim
\mathcal{O}[(\ell_{p}/\mathcal{L})^{n}]$ (where $\delta A =
A_{\gamma} - A_{a}$, $a$ denoting another particle).

As before, we note the presence of the $\pm$ signs which denote
the helicity dependance of the photon. To the order of interest,
equation (\ref{eq:complet E for photons}) can be written
\begin{eqnarray} \label{eq: E for photons}
E^{2}_{\pm} = p^{2} \left[ A_{\gamma}^{2} \pm 2 \theta_{\gamma}
(\ell_{p} p) \right].
\end{eqnarray}
Notably, (\ref{eq: E for photons}) is essentially the same result
that Gambini and Pullin \cite{Gambini Pullin} have obtained for
photon's dispersion relation, with the difference that they have
$A_{\gamma} = 1$ and therefore the semiclassical scale
$\mathcal{L}$ is absent.

A similar contribution was also suggested by Ellis {\it et al}
\cite{Ellis, Ellis2} (in this case, without helicity dependance).
They found
\begin{eqnarray} \label{eq: ellis photons}
E^{2} = p^{2} \left[ 1- 2 M_{D}^{-1} p \right],
\end{eqnarray}
where $M_{D}$ is a mass scale coming from D-brane recoil effects
for the propagation of photons in vacuum. When Gamma Ray Burst
(GRB) data are analyzed to restrict $M_{D}$ \cite{Ellis2}, the
following condition arises
\begin{eqnarray} \label{eq: cond. over MD}
M_{D} \gtrsim 10^{24} eV.
\end{eqnarray}
For the photon's dispersion relation that we are currently
considering, (\ref{eq: cond. over MD}) can be interpreted as the
bound $\theta_{\gamma} \lesssim 10^{4}$. Since $\theta_{\gamma}$
is an adimensional parameter of order one, expression (\ref{eq: E
for photons}) still is a permitted dispersion relation, in what
GRB concern. We shall soon see other possibilities to contrast a
$\theta_{\gamma }$ like term.

\section{Kinematical Approach}

A decay reaction is kinematically allowed when, for a given value
of the total momentum $\vec p_{0} = \sum_{initial} \vec p =
\sum_{final} \vec p$, one can find a total energy value $E_{0}$
such that $E_{0} \geq E_{min}$. Here $E_{min}$ is the minimum
value that the total energy of the decaying products can acquire,
for a given total momentum $\vec p_{0}$. To find $E_{min}$ for the
dispersion relations under consideration, it is enough to take the
individual decay product momenta to be collinear respect to the
total momentum $\vec p_{0}$ and with the same direction. To see
this, it is enough to variate $E_{0}$ with the appropriate
restrictions
\begin{eqnarray}
E_{0} = \sum_{i} E_{i} (|p_{i}|) + \xi_{j} (p_{0}^{\, j} -
\sum_{i} p_{i}^{\, j}),
\end{eqnarray}
where $\xi_{j}$ are Lagrange multipliers, the $i$ index specify
the $i$th particle and the $j$ index the $j$th vectorial component
of the different quantities. Doing the variation, we obtain
\begin{eqnarray}
\frac{\partial E_{i}}{\partial p_{i}^{j}} \equiv v_{i}^{j} =
\xi_{j}.
\end{eqnarray}
That is to say, the velocities of all product particles must be
equal to $\xi$. Since the dispersion relations that we are
treating are monotonously increasing in the range of momenta $p >
\lambda$, this result means that the momenta can be taken
collinear and with the same direction of $\vec p_{0}$.

In the present work, we will focus on those cases in which two
particles (say $a$ and $b$) collide, and lately decay. For the
present, these particles will have momenta $\vec p_{a}$ and $\vec
p_{b}$ respectively, and a total momentum $\vec p_{0}$.
Nevertheless, the total energy of the system will depend only on
$|p_{a}|$ and $|p_{b}|$. Therefore, to get the threshold condition
for the mentioned process, we must find the maximum possible total
energy $E_{max}$ of the initial configuration, given $|p_{a}|$ and
$|p_{b}|$. For this, let us fix $\vec p_{a}$ and variate the
direction of $\vec p_{b} \equiv |p_{b}| \hat n$ in
\begin{eqnarray} \label{eq: restricion 2 over E}
E_{0} & = & E_{a}(\vec p_{0} - |p_{b}| \hat n) + E_{b} (|p_{b}|) +
\chi (\hat n ^{2} - 1).
\end{eqnarray}
Varying (\ref{eq: restricion 2 over E}) respect to $\hat n$
($\chi$ is a Lagrange multiplier), we find
\begin{eqnarray}
\hat n ^{i} = \frac{v_{a}^{i} |p_{b}|}{2 \chi}.
\end{eqnarray}
In this way we obtain two extremal situations $\chi = \pm v_{a}
|p_{b}|/2$, or simply
\begin{eqnarray}
\hat n ^{i} = \pm \frac{v_{a}^{i}}{v_{a}}.
\end{eqnarray}
A simple inspection shows that for the dispersion relations that
we are considering, the maximum energy is given by $\hat n ^{i} =
- v_{a}^{i}/v_{a}$, or in other words, when frontal collision
occurs.

Summarizing, the threshold condition for a two particle ($a$ and
$b$) collision and posterior decay, can be expressed through the
following requirements.
\begin{eqnarray}
E_{a} + E_{b} \geq \sum_{final} E_{f}
\end{eqnarray}
with all final particles having the same velocity, and
\begin{eqnarray}
p_{a} - p_{b}  = \sum_{final} p_{f},
\end{eqnarray}
where the sign of the momenta $\sum_{final} p_{f}$ is given by the
direction of the highest momentum magnitude of the initial
particles. A more detailed treatment can be found in \cite{Coleman
& Glashow}.

As a final remark for this section, under certain circumstances
(for example some special choice of the LID parameters) the
condition $v_{i}^{j} = \xi_{j}$ could give more than one solution
for the threshold-condition configuration. In fact, as noted in
\cite{Liberati}, for a reaction where two identical particles are
the decaying products, it is possible to find configurations where
the momenta of these particles are distributed asymmetrically
within them. However, for the present work, these effects can be
neglected since they give contributions to the threshold
conditions that are smaller than those which we will consider.

\section{Decay Reactions}

Using the methods described in the last section, we can find the
threshold conditions for the decay reactions leading to the
theoretical limits for cosmic rays. These thresholds will present
some consequential modifications due to the parameters of the
theory. Here we examine the possible bounds on these parameters.
Let us start with the observations coming from multi-TeV
$\gamma$-rays.

\subsection{Pair decay, $\gamma + \gamma_{\epsilon} \rightarrow e^{-} + e^{+}$}

Multi-TeV photons are subject to interactions with the FIBR
through the process $\gamma + \gamma_{\epsilon} \rightarrow e^{-}
+ e^{+}$, where $\gamma_{\epsilon}$ is a soft photon from the
FIBR. For this reaction to occur, the following threshold
condition must be satisfied
\begin{eqnarray}
E_{\gamma} + \omega \geq E_{e^{+}} + E_{e^{-}}
\end{eqnarray}
with
\begin{eqnarray}
p_{\gamma} - k = p_{e^{+}} + p_{e^{-}}.
\end{eqnarray}
In the above expressions, $\omega$ and $k$ are the energy and
momentum of the target photon from the FIBR. Since the energy of
these photons do not significantly exceed the eV range, we will
consider for these the usual dispersion relation $w=k$. The above
equations can be reexpressed as
\begin{eqnarray} \label{eq: quadrat E for e+ e-}
E_{\gamma}^{2} + 2 \omega E_{\gamma} \geq E_{e^{+}}^{2} +
E_{e^{-}}^{2} + 2 E_{e^{+}} E_{e^{-}}
\end{eqnarray}
and
\begin{eqnarray}
p_{\gamma}^{2} - 2 k p_{\gamma} = p_{e^{+}}^{2} + p_{e^{-}}^{2} +
2 p_{e^{+}} p_{e^{-}},
\end{eqnarray}
where we have neglected the quadratic terms in the FIBR
quantities. An important property of the field equations from
which the fermion dispersion relation comes from is that they are
charge conjugation invariant. Therefore we can take for both
electron and positron, the same dispersion relation with the same
sign conventions. Furthermore, an analysis of conservation of
angular momenta shows that both helicities are equally probable
for the emerging pair, hence for the right hand of (\ref{eq:
quadrat E for e+ e-}) we must procure that the energy of both,
electron and positron, be the minimum possible. For this reason,
we must use $E_{e^{+}} = E_{e^{-}} = E_{(-)}$, where $E_{(-)}$ is
defined as
\begin{eqnarray}
E_{(-)}^{2} \equiv A^{2} p^{2} + \eta p^{4} - 2 |\lambda| p +
m^{2}.
\end{eqnarray}
Physically, this condition means that the helicity state of less
energy is the one that sets the threshold condition. With this
consideration, we are left with
\begin{eqnarray}
E_{\gamma}^{2} + 2 \omega E_{\gamma} \geq 4 E_{(-)}^{2}.
\end{eqnarray}
>From the dispersion relations (\ref{eq: E for fermions}) and
(\ref{eq: E for photons}) we can write the last equation as
\begin{eqnarray}
p_{\gamma}^{2} [A_{\gamma}^{2} + (\pm)_{\gamma} 2 \theta_{\gamma}
(\ell_{p} p_{\gamma})] + 2 \omega E_{\gamma} \geq 4 [A_{e}^{2}
p_{e}^{2} - 2 |\lambda| p_{e} + m_{e}^{2} ].
\end{eqnarray}
Here, $(\pm)_{\gamma}$ stands for the incident photon helicity.
Note that we have neglected the terms related with $\eta$; these
terms will become important when we study other reactions.
Replacing the momentum conservation, we obtain
\begin{eqnarray} \label{eq: threshold - gamma}
p_{\gamma}^{2} (A_{\gamma}^{2}- A_{e}^{2}) + (\pm)_{\gamma} 2
\theta_{3} \ell_{p} p_{\gamma}^{3} + 2 (\omega E_{\gamma} +
p_{\gamma} k A_{e}^{2}) + 8 |\lambda| p_{e} \geq 4 m_{e}^{2}.
\end{eqnarray}
To the order in consideration we can replace $p$'s by $E$'s.
Additionally, we can use $2 E_{e} \simeq E_{\gamma}$
\begin{eqnarray} \label{eq: threshold for photon - electron}
E_{\gamma}^{2} (A_{\gamma}^{2}- A_{e}^{2}) + (\pm)_{\gamma} 2
\theta_{\gamma} \ell_{p} E_{\gamma}^{3} + 4 \omega E_{\gamma} + 4
|\lambda| E_{\gamma} \geq 4 m_{e}^{2}.
\end{eqnarray}
Now, note that in the absence of quantum gravity corrections we
would have the usual threshold condition
\begin{eqnarray}
E_{\gamma} \geq \frac{m_{e}^{2}}{\omega},
\end{eqnarray}
therefore, to contrast the new terms, we compare them with the
quantity $4 m_{e}^{2}$ in the right side of inequality (\ref{eq:
threshold for photon - electron}).

Following \cite{Stecker & Jager}, no LID should be inferred from
the analysis of the data from the observed Markarian Blazar Mrk
501. This impose strong bounds on our parameters and, in
particular, it means that any modified term must be less than $4
m_{e}^{2}$ up to photons of energy $\sim 20 TeV$. In the first
place, let us see the $A$ terms
\begin{eqnarray}
E_{\gamma}^{2} |A_{\gamma}^{2}- A_{e}^{2}| \cong 2 E_{\gamma}^{2}
|A_{\gamma}- A_{e}| \leq 4 m_{e}^{2}.
\end{eqnarray}
So, it follows that
\begin{eqnarray}
|\delta A| \leq 2 \frac{m_{e}^{2}}{E_{\gamma}^{2}}.
\end{eqnarray}
Evaluating with $E_{\gamma} \sim 20 TeV$, we obtain $|\delta A|
\leq 1.3 \times 10^{-15} $. If we assume that the adimensional
parameters are of order one, and taking for $\Upsilon$ the value
$\Upsilon = -1/2$ (so that $\delta A = \mathcal{O}(
\ell_{p}/\mathcal{L}$)), we can estimate the following bound for
$\mathcal{L}$
\begin{eqnarray}
\mathcal{L}\gtrsim 6.4 \times 10^{-14} eV^{-1}.
\end{eqnarray}
Nevertheless, typical values for the LID parameter difference
$|\delta A|$ are below $10^{-22}$ \cite{Coleman & Glashow}. This
in turn impose a new bound $\mathcal{L} \gtrsim 8.3 \times 10^{-7}
eV^{-1} $ (or $\mathcal{L} \gtrsim 10^{-11} cm$) which is nearly
in the range of nuclear physics. Since there is no evidence that
space manifests its loop structure at this scale, we interpret
this result as that $\delta A = \mathcal{O} (\ell_{p}^{2} /
\mathcal{L}^{2})$ (that is, the universality is broken at most in
second order in the ratio $\ell_{p} / \mathcal{L}$). With this
last assumption we obtain a favoured $\Upsilon = 0$ value, and the
bound
\begin{eqnarray} \label{lobound1}
\mathcal{L}\gtrsim 8.3 \times 10^{-18} eV^{-1}.
\end{eqnarray}
This is by far a more reasonable bound for $\mathcal{L}$.

In the second place, we have the $\theta_{\gamma}$ term (recall
that this term involves the photon helicity dependance). Imposing
the same kind of constrain with photons of energy $E_{\gamma} \sim
20 TeV$, we obtain
\begin{eqnarray} \label{eq: theta-gamma}
|\theta_{\gamma}| \lesssim 0.8.
\end{eqnarray}
This is not a serious bound on the parameter $\theta_{\gamma}$. In
any case, if $|\theta_{\gamma}| \gtrsim 1$ then the observed
photons from Mrk 501 should have a preferred helicity (this
particular helicity will depend on the sign of $\theta_{\gamma}$).
Furthermore, since $\theta_{\gamma}$ is assumed to be a parameter
of order one, expression (\ref{eq: theta-gamma}) tells us that
more energetic photons than those we are considering (energies
$\sim 20 TeV$) should appear with such a preferred helicity.

Finally, it remains the term involving the $\lambda$ parameter for
electrons. For this, we obtain:
\begin{eqnarray}
|\kappa_{5}| \frac{\ell_{p}}{\mathcal{L}^{2}} \leq 2.6 \times
10^{-2} eV.
\end{eqnarray}
Or, assuming that $\kappa_{5}$ is of order one
\begin{eqnarray} \label{lobound2}
\mathcal{L}\gtrsim 5.7 \times 10^{-14} eV^{-1}.
\end{eqnarray}

\subsection{Proton decay, $p + \gamma \rightarrow \Delta$}

The main reaction leading to the GZK limit is the resonant $\Delta
(1232)$ decay $p + \gamma \rightarrow \Delta$. The threshold
condition is
\begin{eqnarray}
E_{p} + \omega \geq E_{\Delta}
\end{eqnarray}
with
\begin{eqnarray}
p_{p} - k = p_{\Delta}.
\end{eqnarray}
Here $E_{\Delta}^{2} = A^{2}p^{2} + \eta p^{4} - 2 |\lambda| p +
m^{2} \,$, that is to say, the minimum possible value for the
energy of the emerging $\Delta$. With some algebraic manipulation
we can find
\begin{eqnarray}
2 \delta A E_{p}^{2} + \delta \eta E_{p}^{4}
 + \left( (\pm)_{p} \, \lambda_{p} +
|\lambda_{\Delta}| \right) E_{p} + 4 \omega E_{p} & \geq &
M_{\Delta}^{2} - M_{p}^{2},
\end{eqnarray}
where $\delta A = A_{p} - A_{\Delta}$ and $\delta \eta = \eta_{p}
- \eta_{\Delta}$. Additionally, $(\pm)_{p}$ refers to the incident
proton helicity. Note that in the absence of LQG modifications the
threshold condition becomes
\begin{eqnarray} \label{eq: GZK class}
E_{p} \geq \frac{M_{\Delta}^{2} - M_{p}^{2}}{4 \omega}.
\end{eqnarray}
Since we don't have a detailed knowledge of the deviation
parameters, we take account of them independently. Naturally,
there will always exist the possibility of having an adequate
combination of these parameter values that could affect the
threshold condition simultaneously. However, as we will soon see,
each one of these parameters will be significant in different
energy ranges.

Let us start considering the terms involving A
\begin{eqnarray}
2 \delta A E_{p}^{2} + 4 \omega E_{p} \geq M_{\Delta}^{2} -
M_{p}^{2}.
\end{eqnarray}
For this inequality it is easy to see that, for a given value of
$\omega$, the reaction is kinematically precluded for all $E$, if
\begin{eqnarray}
A_{\Delta} - A_{p} > \frac{2 \omega^{2}}{M_{\Delta}^{2} -
M_{p}^{2}} \simeq 1.7 \times 10^{-25} \left[ \omega / \omega_{0}
\right] ^{2},
\end{eqnarray}
where $\omega_{0} = 2.35 \times 10^{-4} eV$ is the $kT$ energy
(with $T=2.73 K$) of the CMBR thermal distribution. For all
purposes the GZK limit is forbidden for the CMBR photons if we
take $\omega \simeq \omega_{0}$. Incidentally, assuming that the
adimensional parameters are of order one and that ---as previously
asserted--- the non universal deviation of $A$ is at most of
second order in $\ell_{p}/\mathcal{L}$, we obtain
\begin{eqnarray}
\mathcal{L} \lesssim 2 \times 10^{-16} eV^{-1}.
\end{eqnarray}

Of more relevance than the $A$ terms (as we will verify), are the
$\eta$ related ones. Here we have
\begin{eqnarray} \label{eq: threshold for Delta}
\delta \eta E_{p}^{4} + 4 \omega E_{p} & \geq & M_{\Delta}^{2} -
M_{p}^{2}.
\end{eqnarray}
In this case the condition is independent of $\mathcal{L}$ and it
depends strictly on the difference $\delta \eta$. For this, the
reaction is forbidden if
\begin{eqnarray} \label{eq: eta3}
(\eta_{\Delta} - \eta_{p}) > \frac{27 \omega^{4}}{(M_{\Delta}^{2}
- M_{p}^{2})^{3}} \simeq 3.2 \times 10^{-67}
[\omega/\omega_{0}]^{4} \, eV^{-2}.
\end{eqnarray}
Recalling that $\eta = \kappa_{3} \ell_{p}^{2} \,$, (\ref{eq:
eta3}) tell us that it is enough to have $|\kappa_{3}| > 5 \times
10^{-11}$ with $\kappa_{3 \Delta} > \kappa_{3 p}$, for the
reaction to be precluded. Since we are assuming that
$\mathcal{O}(\kappa_{3})=1$, this result reveals us that the
presence of a non null $\delta \eta < 0$ ensures the GZK violation
effect.

In view of the possibilities $\delta \eta = 0$ and $\delta A = 0$,
we must consider the $\lambda$ dependant terms
\begin{eqnarray}
2 \left( (\pm)_{p} \lambda_{p} + |\lambda_{\Delta}| \right) E_{p}
+ 4 \omega E_{p} & \geq &  M_{\Delta}^{2} - M_{p}^{2}.
\end{eqnarray}
This last expression is very interesting faced with the fact that
its terms are helicity dependant. In this case, the reaction is
more sensitive to the energy of the target photon. For instance,
if $\omega$ is such that
\begin{eqnarray}
(\pm)_{p} \, \lambda_{p} +  |\lambda_{\Delta}| + 2 \omega \leq 0,
\end{eqnarray}
the reaction would be forbidden. Of course, this situation will
depend on the helicity configuration of the incident proton. For
example, if
\begin{eqnarray} \label{lambda conditon}
|\lambda_{p}| \geq  |\lambda_{\Delta}| + 4.7 \times 10^{-4}
[\omega/\omega_{0}] \, eV,
\end{eqnarray}
the reaction would also be forbidden at least for one proton
helicity. Indeed, if $|\kappa_{5 p}| - |\kappa_{5 \Delta}| \gtrsim
1$ (recall that $\lambda = \kappa_{5} \ell_{p} / 2
\mathcal{L}^{2}$) the threshold condition is dominated by the
$\lambda_{p}$ term
\begin{eqnarray} \label{eq: thresh for helicity proton}
|\lambda_{p}| \gtrsim 4.7 \times 10^{-4} [\omega/\omega_{0}] \,
eV.
\end{eqnarray}
This impose a new bound on the parameters of the theory
\begin{eqnarray} \label{upbound2}
\mathcal{L} \lesssim 3 \times 10^{-13} eV^{-1}.
\end{eqnarray}
On the other hand, if $|\kappa_{5 \Delta}| - |\kappa_{5 p}|
\gtrsim 1$, the conservation of angular momentum always allows the
reaction, and no GZK violation would be obtained. Although for
this effect to be notorious we must demand universality on both
$A$ and $\eta$ (at least for these hadronic particles). Since
$\Delta$'s and protons have different spins, we cannot discard
this possibility.

\subsection{Photo - Pion Production, $p + \gamma \rightarrow p + \pi$}

The next relevant reaction leading to the GZK threshold is the
non-resonant photo-pion production $p + \gamma \rightarrow p +
\pi$. Since the pion is a spin 0 particle, we may assume that, to
the order considered for (\ref{eq: E for fermions}), the relevant
dispersion relation is
\begin{eqnarray}
E^{2} = A_{\pi}^{2} p^{2} + \eta_{\pi} \, p^{4} + m_{\pi}^{2},
\end{eqnarray}
where $A_{\pi} = 1 + \kappa_{\pi} (\ell_{p}^{2} /
\mathcal{L}^{2})$ (recall that we must have $\delta A \sim
\ell_{p}^{2} / \mathcal{L}^{2}$). As in the other cases, the
threshold condition will be given by
\begin{eqnarray} \label{cons of energy p-pi}
E_{p} + \omega \geq \bar E_{p} + E_{\pi}
\end{eqnarray}
with
\begin{eqnarray}
p_{p} - k = \bar p_{p} + p_{\pi}.
\end{eqnarray}
Where $\bar E_{p}$ and $\bar p$ refer to the emerging proton. In
analogy with the $\Delta$ decay, for this threshold condition we
must put $\bar E_{p}^{2} = A_{p}^{2} \bar p^{2} + \eta \bar p^{4}
- 2 |\lambda_{p}| \bar p + m_{p}^{2} \,$.

With a little amount of algebra we are able to find
\begin{eqnarray}
2 \delta A E_{\pi}^{2} + \left( \delta \eta + 3 \eta_{p}
\frac{M_{p} ( M_{p}+M_{\pi})}{M_{\pi}^{2}} \right) E_{\pi}^{4} +
4E_{\pi} \omega + 2 E_{\pi} (|\lambda| \pm \lambda) \geq
\frac{M_{\pi}^{2}(2 M_{p}+M_{\pi})}{M_{p}+M_{\pi}},
\end{eqnarray}
where $\delta A = A_{p} - A_{\pi}$, and $\delta \eta = \eta_{p} -
\eta_{\pi}$. In the last expression, $\pm$ refers to the helicity
of the incoming proton. Since there will necessarily be an
incident proton helicity that can minimize this term, we can take
for the threshold condition
\begin{eqnarray}
2 E_{\pi} (|\lambda| \pm \lambda) = 0.
\end{eqnarray}
With this consideration in mind, we get
\begin{eqnarray}
2 \delta A E_{\pi}^{2} + \left( \delta \eta + 168 \eta_{p} \right)
E_{\pi}^{4} + 4E_{\pi} \omega \geq \frac{M_{\pi}^{2}(2
M_{p}+M_{\pi})}{M_{p}+M_{\pi}}.
\end{eqnarray}

As before, let us consider the modifications separately. If
$\delta A$ were the dominant term, we would have to consider
\begin{eqnarray}
2 \delta A E_{\pi}^{2} + 4E_{\pi} \omega \geq \frac{M_{\pi}^{2}(2
M_{p}+M_{\pi})}{M_{p}+M_{\pi}},
\end{eqnarray}
consequently, the violation condition would be
\begin{eqnarray}
A_{\pi} - A_{p} > \frac{2 \omega^{2}
(M_{p}+M_{\pi})}{M_{\pi}^{2}(2 M_{p}+M_{\pi})} \simeq 3.3 \times
10^{-24} \left[ \omega / \omega_{0} \right] ^{2}.
\end{eqnarray}
Using $\delta A \sim \ell_{p}^{2} / \mathcal{L}^{2}$, this result
can be understood as
\begin{eqnarray} \label{upbound1}
\mathcal{L} \lesssim 4.6 \times 10^{-17} eV^{-1}.
\end{eqnarray}

Let us now consider the $\eta$ terms. For these we have a violated
threshold if
\begin{eqnarray}
-\delta \eta - 168\eta_{p} > 27 \omega^{4} \left(
\frac{M_{p}+M_{\pi}}{M_{\pi}^{2}(2 M_{p}+M_{\pi})} \right)^{3}
\simeq 2.2 \times 10^{-63} [\omega/\omega_{0}]^{4} \, eV^{-2}.
\end{eqnarray}
Since $\mathcal{O}(\delta \eta) \simeq \mathcal{O}(\eta_{p})$
(when $\delta \eta \neq 0$), let us assume that the $\eta_{p}$
term dominates. In this case, for the threshold condition to be
violated we just require $|\eta| > 1.3 \times 10^{-65} eV^{-2}$
with $\eta$ negative. Recalling that $\eta = \kappa_{3}
\ell_{p}^{2}$ with $\kappa_{3}$ of order one, this condition can
be well read as $|\kappa_{3}| \geq 1.9 \times 10^{-9}$. Hence if
$\kappa_{3}$ is not strictly zero, this term would be an
acceptable causing agent of the GZK limit violation, as far as
photo-pion production is concerned. Finally, if $\eta$ were null,
the next relevant terms would be the $\lambda$ helicity dependant
ones. But, by angular momentum conservation, there will always be
a emergent proton helicity that cancels them, hence these terms
cannot forbid the reaction.

\section{Conclusions}

We have seen how the introduction of modifications from loop
quantum gravity can affect and explain the anomalies observed in
highly energetic phenomena like cosmic rays. In particular, the
notable appearance of helicity dependant decays could be a special
footprint of this kind of effective theories.

Provided that the difference $\delta A$ between $A_{\gamma}$ and
$A_{e}$ is not affecting the observations of the arrival of
multi-TeV photons (as the ultimate analysis show), we have the
strong possibility granted by (\ref{eq: E for photons})
\begin{eqnarray} \nonumber
E^{2}_{\pm} = p^{2} \left[ A_{\gamma}^{2} \pm 2 \theta_{\gamma}
(\ell_{p} p) \right],
\end{eqnarray}
to be on the edge of observing polarized multi-TeV photons. In a
few words, the actual universe could be transparent for one
helicity state (while for the other not), nearly over the $TeV$
range. The specific helicity necessarily depends on the sign of
$\theta_{\gamma}$ and for the moment no related observations could
decide this sign.

Likewise, there also is the possibility that we have been
observing polarized protons in the form of GZK limit violating
events. For these helicity effects to take place, it is necessary
that both $A$ and $\eta$ be universals parameters as opposed to
$\lambda$, which would need to respect (\ref{lambda conditon}).
This last assumption appears to be a little forced. Nevertheless,
faced with the fact that these terms depend on the parity and CPT
violation structure of the theory and hence the helicity
degeneracy of states is broken, we must take this possibility
seriously. For instance, it is enough to note the great difference
that, in what these helicity terms refers, must exist between
particles of spin zero and fractional spin.

Summarizing, the GZK limit can be violated by $A$, $\eta$ and
$\lambda$ in three different ways. Firstly, by having a non
universal $A$ parameter up to second order in $\ell_{p} /
\mathcal{L}$ ($\Upsilon = 0$ in the case of photons) in such a way
that $A_{p} < A_{\Delta}$ and $A_{p} < A_{\pi}$. In this case,
from bounds (\ref{lobound1}) and (\ref{upbound1}), the favoured
range for $\mathcal{L}$ is
\begin{eqnarray} \label{eq: lenght range 1}
4.6 \times 10^{-17} eV^{-1} \gtrsim \, \mathcal{L} \, \gtrsim \,
8.3 \times 10^{-18} eV^{-1}.
\end{eqnarray}
Note however that this possibility necessarily excludes the
existence of a $\lambda$ term in the dispersion relations for
fermions, since there would be fermions having velocities
in the opposite direction from that of the momentum
(up to $ p= \lambda \simeq KeV$). This last assumption has,
at the same time, the consequence that no parity violation (and
therefore CPT violation) should be present in the fermionic part
of the theory, at the discussed level.

Secondly, by having $\eta_{p} < \eta_{\Delta}$ with $\eta$ a
negative parameter. This case is more interesting since it fixes
the sign of $\eta$ and, consequently, its effects could be studied
in other high energy reactions with at least a little more
knowledge on these corrections.

Thirdly, the already mentioned possibility of an helicity
dependant violation of the limit needs, as in the previous case, a
negative and universal $\eta$. The reason for this exotic
combinations of parameters is that for the photo-pion production
to be forbidden it is only necessary to have a negative $\eta$,
while for the resonant $\Delta$ decay, the $\eta$ sign is not
sufficient. For this last effect to take place, the length scale
$\mathcal{L}$ needs to satisfy (from bounds (\ref{lobound2}) and
(\ref{upbound2}))
\begin{eqnarray} \label{eq: lenght range 2}
3 \times 10^{-13} eV^{-1} \gtrsim \, \mathcal{L} \, \gtrsim \, 5.7
\times 10^{-14} eV^{-1}.
\end{eqnarray}
It is worth noting that the helicity dependant effects tend to
grant a privilege to a length scale around $\sim 2 \times 10^{-13}
eV^{-1}$ or, if we prefer, a mass scale in the $TeV$ range. This
is the same tentative range found in other works related with
gravity \cite{Cohen Kaplan Nelson}. For example, recent works on
compactification of extra dimensions \cite{Arkani-Hamed Dimopoulos
Dvali, Antoniadis Arkani-Hamed Dimopoulos Dvali} show the
possibility to define a mass scale in the $TeV$ range and, as
commonly emphasized, it is on the edge of actual empirical
observations \cite{Dimopoulos Landsberg}. A length scale's range
like (\ref{eq: lenght range 2}) gives a $\lambda$ value
\begin{eqnarray}
\lambda \sim 2.5 \times 10^{-3} eV.
\end{eqnarray}
As was noted in \cite{Carmona Cortes}, a dispersion relation of
the type
\begin{eqnarray}
E^{2}=p^{2} + \lambda p + m^{2},
\end{eqnarray}
with a value of $\lambda \geq 10^{-7} eV$, should be discarded
because of the extremely sensitive measures made on the Lamb
shift. However in the present framework, since the Lamb shift
depends primarily on the interaction detail between electrons and
photons, we are compelled to wait for a complete interaction
picture of the effective LQG theories to say something about the
symmetries involved in a low energy effect like that. In this
sense, our developments are strictly valid for analysis made on
asymptotically free particle states (as done in the present work),
where the effects of interactions are taken to be negligible, and
kinematical considerations are valid.

Future experimental developments like the Auger array, the Extreme
Universe Space Observatory (EUSO) and Orbiting Wide-Angle Light
Collectors (OWL) satellite detectors, shall increase the precision
and phenomenological description (such as a favoured proton
helicity) of these UHECR.

Other related bounds for these parameters can also be established.
Such is the case of Gamma Ray Burst (GRB) observations which could
throw sensitive results on the $\delta A$ difference between
photons and neutrinos \cite{Neutrinos} (see Appendix \ref{Appendix
A}), or neutrino oscillations, in which the universality between
the different neutrino flavor LID parameters can be measured
\cite{Oscillations}.

{\acknowledgments The authors are grateful to S. Liberati and A.
Ringwald for calling to our attention references \cite{Liberati} and
\cite{Z-Burst1, Z-Burst3}. Also  to S. G\'{a}lvez for
his help in improving the English of the paper. The work of
JA is partially supported by Fondecyt 1010967. The work of GP is
partially supported by a CONICYT Fellowship.}

\appendix
\section{Time Delay Between Photons and Neutrinos from GRB} \label{Appendix A}

The prediction of $10^{14}-10^{19} eV$ neutrino bursts generated
in GRB events \cite{Waxman & Bahcall, Vietri}, open the
interesting possibility of observing a time delay between the
arrival of photons and neutrinos. For instance, taking into
account the range predicted in (\ref{eq: lenght range 1}) ---which
gives an $A$ difference of $\delta A \approx 10^{-22}$---, the
time delay from a typical source at 40 Mpc in a flat
Friedman-Robertson-Walker (FRW) universe, should be $\delta t \sim
10^{-6}s$. This result may be compared with the respective one
from \cite{Neutrinos}, where, for the same distance, it is found
that $\delta t \sim 0.4 \times 10^{2}s$. The great discrepancy can
be understood not only on the ground of having different magnitude
and expressions for $\delta A$ in terms of the $\mathcal{L}$
parameter, but also on the fact that in \cite{Neutrinos} the
length scale $\mathcal{L}$ was taken to be a mobile scale which
sets a cutoff value for the involved momenta ($\mathcal{L} \lesssim
1/p$) given the specific physical situation. In this paper we have
considered that $\mathcal{L}$ is a universal length scale. From
this point of view, since the length scale is not mobile, we have
included the possibility $p > 1/ \mathcal{L}$ (in which the LQG
structure of the region $\ell_{p} \lesssim d \lesssim \mathcal{L}$
is present through its effects), and therefore the actual results.

For completeness, let us show the time delay contributions (in a
flat FRW universe) from the most significant terms of the
dispersion relation for neutrinos. These delays are considered
respect to the arrival of photons with a conventionally re-scaled
$A_{\gamma} \equiv 1$.
\begin{itemize}
\item{$A$ term:}
\begin{eqnarray} \label{A1}
\delta t_{A} = \frac{2 |\delta A|}{H_{0}}[1-(1+z)^{-1/2}].
\end{eqnarray}
\item{$\eta_{\nu}$ term:}
\begin{eqnarray} \label{A2}
\delta t_{\eta} = \frac{|\eta_{\nu}|
p_{0}^{2}}{H_{0}}[(1+z)^{3/2}-1].
\end{eqnarray}
\end{itemize}

Additionally, there will be a time delay between photons of
different helicities \cite{Photons} (to follow the later
convention, we take $(v_{+} + v_{-})/2 = A_{\gamma} \equiv 1$,
where $v_{\pm} = A_{\gamma} \pm 2 \theta_{\gamma} \ell_{p} p$)
\begin{itemize}
\item{$\theta_{\gamma}$ term:}
\begin{eqnarray} \label{A3}
\delta t_{\pm} = \frac{8 |\theta_{\gamma}| \ell_{p}
p_{0}}{H_{0}}[(1+z)^{1/2}-1].
\end{eqnarray}
\end{itemize}
In (\ref{A1}), (\ref{A2}) and (\ref{A3}), $p_{0}$ is the momentum
(or energy) of the arriving particles, $H_{0}$ is the Hubble
constant and $z$ is the source redshift. The above results can be
used to analyze the GRB spectral structure in more detail and give
additional bounds to the current parameters. Currently, present
observations can not give such bounds.

\end{document}